# Algorithm-Hardware Co-Optimization of the Memristor-Based Framework for Solving SOCP and Homogeneous QCQP Problems

Ao Ren[1], Sijia Liu[2], Ruizhe Cai[1], Wujie Wen[3], Pramod K. Varshney[1], Yanzhi Wang[1]

[1]Syracuse University, [2]University of Michigan, [3]Florida International University, U.S.A

[1]{aren, rcai100, varshney, ywang393}@syr.edu, [2]lsjxjtu@umich.edu, [3]wwen@fiu.edu

**Abstract** - **A memristor crossbar, which is constructed with memristor devices, has the unique ability to change and memorize the state of each of its memristor elements. It also has other highly desirable features such as high density, low power operation and excellent scalability. Hence the memristor crossbar technology can potentially be utilized for developing low-complexity and high-scalability solution frameworks for solving a large class of convex optimization problems, which involve extensive matrix operations and have critical applications in multiple disciplines. This paper, as the first attempt towards this direction, proposes a novel memristor crossbar-based framework for solving two important convex optimization problems, i.e., second-order cone programming (SOCP) and homogeneous quadratically constrained quadratic programming (QCQP) problems. In this paper, the alternating direction method of multipliers (ADMM) is adopted. It splits the SOCP and homogeneous QCQP problems into sub-problems that involve the solution of linear systems, which could be effectively solved using the memristor crossbar in O(1) time complexity. The proposed algorithm is an iterative procedure that iterates a constant number of times. Therefore, algorithms to solve SOCP and homogeneous QCQP problems have pseudo-O($N$) complexity, which is a significant reduction compared to the state-of-the-art software solvers (O($N^{3.5}$) - O($N^4$)).**

## I. Introduction

Convex optimization is a research field that aims to find the optimal solution for the problem of minimizing a convex objective function subject to some convex constraints. The utility of convex optimization has been shown extensively in various applications such as signal processing, communications, smart grid, machine learning, circuit design, and other applications [1][2]. It is especially required in state-of-the-art large-scale applications in machine learning (e.g., the support vector machine [3]) and compressed sensing techniques [4].

There does not exist a common solution for general convex optimization problems [5]. But for a number of important types of convex optimization problems, such as semidefinite programming (SDP), quadratically constrained quadratic programming (QCQP) and second-order cone programming (SOCP), optimal software-based solutions exist that use effective algorithms such as extensions of the primal-dual interior point (PDIP) method [8]. However, in the era of data deluge, software-based optimization solvers suffer from limited scalability in high-dimensional data regimes. For example, solving a SDP problem has an O($N^6$) complexity using state-of-the-art software-based solvers [20]. This complexity is prohibitive for problems with large volumes of data. Therefore, it is imperative to develop new techniques and new solvers that overcome these limitations.

The recently invented memristor crossbar can potentially resolve the limitations efficiently. Because the memristor device, invented by HP Lab [13], has the unique property that its state (memristance) can be changed when the voltage drop at its two terminals is higher than a threshold voltage. Thus, a single memristor device can be readily utilized to represent a matrix element. Moreover, its promising features of non-volatility, excellent scalability, high density and low power operation make it a candidate to be arranged in a crossbar structure to represent matrices and perform matrix computations efficiently (often in O(1) time complexity). As many convex optimization problems, such as SOCP problems, need to perform a large number of matrix operations (matrix-vector multiplications and solving linear systems, etc.), they can potentially be solved by using memristor crossbar technology that provides low computational complexity, high speed and energy efficiency.

Despite the fact that memristor devices have the potential to be utilized to solve certain important convex optimization problems, there are multiple challenges and limitations from both algorithm side and hardware side. From the algorithm side, the algorithms proven to be successful in solving SOCP problems with software-based solvers may not be appropriate for hardware implementations. With respect to the hardware side, the memristor crossbar can only deal with square matrix computations and the matrix elements can only be non-negative numbers because memristance cannot be negative. Consequently, an algorithm-hardware co-design and co-optimization framework is required to overcome these limitations with high efficiency and low computational complexity.

For ease of hardware implementation, we use an operator splitting method, the alternating direction method of multipliers (ADMM), to solve SOCP problems. The major advantage of ADMM is that it can split the original problem into a set of problems that involve the solution of linear systems. Additionally, a large number of problems can be formulated in the form of SOCP or be formulated as problems with second-order cone constraints, such as homogeneous QCQP problems [5]. Hence, a large number of convex optimization problems can be solved efficiently with the memristor crossbar and ADMM algorithm.

To the best of our knowledge, this paper presents the first framework for solving SOCP and homogeneous QCQP problems using memristor crossbar techniques. This is expected to be an important step towards the solution of more general convex optimization problems. The proposed solution procedure is an iterative procedure with O($N$) complexity in each iteration, and the procedure does not need to update the conductance matrix of memristor crossbar during iterations, thereby significantly reducing the solution complexity. Besides, the procedure only iterates a constant number of times, thus the solution framework can achieve pseudo-O($N$) computational complexity. Compared with software-based solvers of SOCP and homogeneous QCQP problems (the CVX tool), the proposed memristor crossbar-based solution framework achieves significant speedup and energy efficiency improvement up to $1.57 \times 10^5$X and $1.32 \times 10^7$X, respectively. Finally, extensive experimental results demonstrate excellent reliability of the proposed solution framework under process variations.

In the rest of this paper, Section II presents the background on convex optimization and the forms of SOCP and homogeneous QCQP problems, as well as the memristor crossbar structure and its properties. Section III describes our memristor-based framework and the procedures to solve the SOCP and homogeneous QCQP problems using it. Section IV analyzes and explains our experimental results. Conclusion is provided in Section V.

*This research is sponsored in part by a grant from the Software and Hardware Foundations of the Division of Computer and Communication Foundations of the U.S. National Science Foundation.

## II. Background

### A. Convex Optimization

Convex optimization arises in a variety of applications, such as automatic control, communications, signal processing, and the state-of-the-art large-scale applications in machine learning and compressed sensing techniques [3][4]. Convex optimization is attractive since a local optimum is also a global optimum in convex programs and a rigorous optimality condition and duality theory exist to verify the optimal solution. Its standard form consists of three parts: an objective function which must be a convex function, a set of inequality constraints which must be convex as well, and a set of equality constraints which must remain affine. A convex function can be written as:

$$f_i(\theta x + (1-\theta)y) \leq \theta f_i(x) + (1-\theta)f_i(y) \quad (1)$$

where $0 \leq \theta \leq 1$ [5]. Therefore, a convex minimization problem is written as:

$$\begin{array}{ll} \text{minimize} & f_0(\pmb{x}) \\ \text{subject to:} & f_i(\pmb{x}) \leq 0 \ (i=1,\ldots,m), \\ & h_i(\pmb{x}) = 0 \ (i=1,\ldots,p) \end{array} \quad (2)$$

where the optimization variables $\pmb{x} \in \mathbb{R}^n$, and $f_0, f_1, \ldots, f_m$ are convex functions: $\mathbb{R}^n \rightarrow \mathbb{R}$ [5].

There is no general polynomial-time solution for the most general type of convex optimization problems [5], but many types of convex optimization problems, including QCQP, SOCP, and SDP, can be solved in polynomial-time complexity (generally $O(N^{4.5})$ to $O(N^{6.5})$) [6][7] using carefully designed algorithms, e.g., variants of the PDIP method [8].

### B. Second-Order Cone Programming (SOCP)

SOCP is a convex program to minimize a linear function over a set of linear constraints and the product of second-order cones [9]. It has wide applications in resource allocation in wireless communication networks, high-performance computing, smart grid, etc. [10]-[12]. For example, coordinated beamforming in wireless communication systems [24] yielded a direct SOCP formulation given as:

$$\begin{array}{ll} \text{minimize} & \pmb{c}^T \pmb{x} \\ \text{subject to:} & \pmb{Ax} = \pmb{b}, \\ & \|\pmb{x}_{1:(n-1)}\|_2 \leq x_n, \end{array} \quad (3)$$

where $\pmb{x}$ is the optimization variable, $\pmb{x}_{1:(n-1)}$ is the vector that consists of the first ($n$-1) entries of $\pmb{x}$, $x_n$ is the $n$-th entry of $\pmb{x}$. The last constraint represents a second-order cone in $\mathbb{R}^n$.

### C. Homogeneous Quadratically Constrained Quadratic Programming (QCQP)

If the objective function and the inequality constraints are convex quadratic, then it is called a quadratically constrained quadratic problem (QCQP). A QCQP is homogeneous if all quadratic functions do not have any linear terms. Homogenous QCQPs were commonly used to solve the problems of resource management in signal processing, such as optimal power allocation for linear coherent estimation [25] and optimal spectrum sharing in MIMO cognitive radio networks [26]. A homogeneous QCQP problem has the form:

$$\begin{array}{ll} \text{minimize} & \pmb{x}^T \pmb{P}_0 \pmb{x} \\ \text{subject to:} & \pmb{x}^T \pmb{P}_i \pmb{x} \leq \pmb{r}_i \ (i=1,\ldots,m), \\ & \pmb{Ax} = \pmb{b}, \end{array} \quad (4)$$

which can be converted to an SOCP problem. Hence in this paper, we mainly focus on the solution of SOCP problems.

### D. Memristor Crossbar

The memristor device was invented by HP lab in 2008 [13]. The most important feature of the memristor device is its unique ability to record the historical profile of the excitations on the device. More specifically, when the voltage applied at its two terminals is higher than a threshold voltage, i.e., $|V_m| > |V_{th}|$, the state (memristance) of a memristor will change. Otherwise, the memristor behaves like a resistor. Thus, it is attractive to use memristors for matrix computations (often with O(1) time complexity) because a memristor can be used to represent an element of a matrix. In addition, it has many other promising features, such as non-volatility, low-power operation, high density, and excellent scalability [13][14]. Hence, the challenges for software-based convex optimization solvers, such as limited scalability, excessive overhead of time and energy consumptions in high dimensional data regimes, can be potentially resolved by properly using memristor devices.

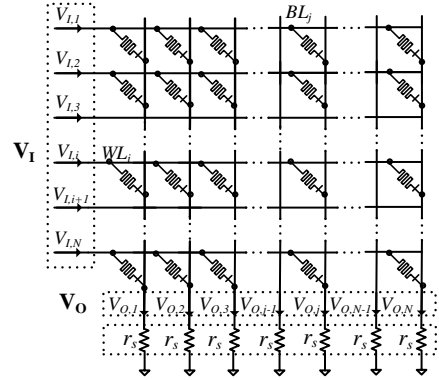

Fig. 1. A typical $N \times N$ memristor crossbar

A typical $N \times N$ memristor crossbar is illustrated in Fig. 1, in which a memristor is connected between each pair of horizontal word-line (WL) and vertical bit-line (BL). By properly applying biasing voltages at its two terminals, each memristor can be re-programmed to different resistance states so that the memristor crossbar can be utilized to represent matrices [15]-[17]. To demonstrate the matrix computation functionality, we apply a vector of input voltages $\pmb{V_I}$ on WLs and collect the current through each BL by measuring the voltage across resistor $r_s$ with conductance of $g_s$. Assume that the memristor that connects $WL_i$ and $BL_j$ has a conductance of $g_{i,j}$, then the output voltages can be represented by $\pmb{V_O} = \pmb{C} \times \pmb{V_I}$. $\pmb{C}$ is determined by the conductance of memristors as follows:

$$\pmb{C} = \pmb{D} \cdot \pmb{G}^T = diag(d_1,\ldots,d_N) \cdot \begin{bmatrix} g_{1,1} & \cdots & g_{1,N} \\ \vdots & \ddots & \vdots \\ g_{N,1} & \cdots & g_{N,N} \end{bmatrix}^T \quad (5)$$

where $d_i = 1/(g_s + \sum_{k=1}^{N} g_{k,i})$. Accordingly, a matrix multiplication is conducted by the memristor crossbar with the time complexity of O(1).

In the reverse direction, the memristor crossbar structure can solve a linear system of equations [18]. By mapping the coefficient matrix of a set of linear equations, and applying a voltage vector $\pmb{V_O}$ on each $r_s$ of BLs, the current flowing through each BL can be approximated as $I_{o,j} = g_s V_{o,j}$. On the other hand, current $I_{o,j}$ through $BL_j$ can also be calculated as $I_{o,j} = \sum_j V_{I,i} g_{i,j}$. Hence, for each $BL_j$, equation $\frac{1}{g_s}\sum_j V_{I,i} g_{i,j} = V_{o,j}$ is mapped and the solution $\pmb{V_I}$ can be determined by measuring voltages on the WLs. Thus, the system of linear equations $\pmb{C} \cdot \pmb{V_I} = \pmb{V_O}$ is mapped to the memristor crossbar structure.

Note that since matrix coefficients are represented by memristance values in a memristor crossbar, only non-negative coefficients can be supported when solving a linear system. For matrix-vector multiplication, this limitation can be mitigated by using two matrices, one that stores positive values $\pmb{C_+}$ and the other that stores negative values $\pmb{C_-}$, and performing subtraction $\pmb{C_+} \times \pmb{V_I} - \pmb{C_-} \times \pmb{V_I}$ using summing amplifiers [16][17].

## III. Memristor Crossbar-Based Framework For Solving Convex Optimization Problems

### A. Alternating Direction Method of Multipliers (ADMM)

It has been recently shown that ADMM is a powerful tool for solving large-scale optimization problems; examples include sensor scheduling in large networks of dynamic systems [27], resource allocation in dense wireless cooperative networks [12] and design of feedback control systems [28]. The major advantage of ADMM is that it allows us to split the original problem into sub-problems, each of which can be solved more efficiently or even analytically. ADMM solves convex problems of the form [19]:

$$\text{minimize } f(\mathbf{x}) + g(\mathbf{y}) \text{ subject to: } \mathbf{x} = \mathbf{y} \quad (6)$$

where $f$ and $g$ may be non-smooth or take infinite values to encode implicit constraints.

ADMM is an iterative method. Its $k$-th iteration is:

$$\mathbf{x}^{(k+1)} = \arg\min_{\mathbf{x}}(f(\mathbf{x}) + g(\mathbf{y}) + (\rho/2)||\mathbf{x} - \mathbf{y}^{(k)} + (1/\rho)\boldsymbol{\mu}^{(k)}||_2^2) \quad (7a)$$

$$\mathbf{x}^{(k+1)} = \arg\min_{\mathbf{x}}(f(\mathbf{x}) + g(\mathbf{y}) + (\rho/2)||\mathbf{x} - \mathbf{y}^{(k)} + (1/\rho)\boldsymbol{\mu}^{(k)}||_2^2) \quad (7b)$$

$$\boldsymbol{\mu}^{(k+1)} = \boldsymbol{\mu}^{(k)} + \rho(\mathbf{x}^{(k+1)} - \mathbf{y}^{(k+1)}) \quad (7c)$$

where $\rho > 0$ is the step size parameter, and $\boldsymbol{\mu}$ is the dual variable associated with the constraint $\mathbf{x} = \mathbf{y}$. Under some mild conditions [19], ADMM converges to the optimal solution of problem (6).

### B. Solving SOCP Using Memristor Crossbar via ADMM

In order to solve a SOCP problem via the use of ADMM, we reformulate the problem in (6) by introducing a new variable $\mathbf{y} \in \mathbb{R}^n$ and incorporating the indicator function, then the original problem is equivalent to

$$\begin{aligned} \text{minimize} \quad & \mathbf{c}^T\mathbf{x} + \mathbf{I}_1(\mathbf{x}) + \mathbf{I}_2(\mathbf{y}) \\ \text{subject to:} \quad & \mathbf{x} = \mathbf{y} \end{aligned} \quad (8)$$

where $\mathbf{x}$ and $\mathbf{y}$ are optimization variables, and $\mathbf{I}_1$ and $\mathbf{I}_2$ are indicator functions given by

$$\mathbf{I}_1(\mathbf{x}) = \begin{cases} 0, & \text{if } \mathbf{A}\mathbf{x} = \mathbf{b} \\ \infty, & \text{otherwise} \end{cases} \quad (9a)$$

$$\text{and } \mathbf{I}_2(\mathbf{y}) = \begin{cases} 0, & \text{if } \|\mathbf{y}_{1:(n-1)}\|_2 \leq y_n \\ \infty, & \text{otherwise} \end{cases} \quad (9b)$$

A direct application of ADMM to SOCP yields the x-minimization step and y-minimization step. That is:

- x-minimization step:

$$\begin{aligned} \text{minimize} \quad & \left(\frac{1}{2}\right)\mathbf{x}^T\mathbf{x} - \left(\mathbf{u}^{(k)}\right)^T\mathbf{x} \\ \text{subject to:} \quad & \mathbf{A}\mathbf{x} = \mathbf{b} \end{aligned} \quad (10)$$

where $\mathbf{u}^{(k)} \equiv \mathbf{y}^{(k)} - (1/\rho)(\boldsymbol{\mu}^{(k)} + \mathbf{c})$, and $\boldsymbol{\mu}^{(k)}$ can be calculated by Eqn. (7c).

- y-minimization step:

$$\begin{aligned} \text{minimize} \quad & \|\mathbf{y} - \mathbf{v}^{(k)}\|_2^2 \\ \text{subject to:} \quad & \|\mathbf{y}_{1:(n-1)}\|_2 \leq y_n \end{aligned} \quad (11)$$

where $\mathbf{v}^{(k)} \equiv \mathbf{x}^{(k+1)} + (1/\rho)\boldsymbol{\mu}^{(k)}$.

The optimal solution of the x-minimization problem is readily obtained by using the Lagrangian method. By introducing a multiplier $\boldsymbol{\lambda} \in \mathbb{R}^m$, the Lagrangian of (10) can be written as

$$\mathcal{L}(\mathbf{x}, \boldsymbol{\lambda}) = \left(\frac{1}{2}\right)\mathbf{x}^T\mathbf{x} - \left(\mathbf{u}^{(k)}\right)^T\mathbf{x} + \boldsymbol{\lambda}^T(\mathbf{A}\mathbf{x} - \mathbf{b}) \quad (12)$$

Taking the first derivatives of $\mathcal{L}(\mathbf{x}, \boldsymbol{\lambda})$ with respect to $\mathbf{x}$ and $\boldsymbol{\lambda}$ and setting them equal to zero yields the following linear system of equations:

$$\begin{bmatrix} \mathbf{I}_{n \times n} & \mathbf{A}^T \\ \mathbf{A} & \mathbf{0}_{m \times m} \end{bmatrix} \begin{bmatrix} \mathbf{x} \\ \boldsymbol{\lambda} \end{bmatrix} = \begin{bmatrix} \mathbf{u}^{(k)} \\ \mathbf{b} \end{bmatrix} \quad (13)$$

A memristor crossbar could be utilized to effectively solve the linear system of equations and derive $\mathbf{x}$ and $\boldsymbol{\lambda}$ by configuring memristance values according to the left-hand side matrix and applying the right-hand side vector at the output end of the memristor crossbar. The left-hand side matrix is already a square matrix and therefore suitable for memristor crossbar-based implementations.

Because negative elements may exist in matrix $\mathbf{A}$, a special treatment using additional variables is required to eliminate the negative coefficients and maintain a square matrix. Details of the method for mapping the left-hand side matrix onto a memristor crossbar and the method for dealing with negative elements are covered in Section C of this paper.

The optimal solution of the y-minimization problem given in (11) can be obtained in a closed form and given by projecting a point $\mathbf{v}^{(k)}$:

$$\mathbf{v}^{(k)} = \left[\left(\mathbf{w}^{(k)}\right)^T, s\right]^T \quad (14)$$

onto a second-order cone in $\mathbb{R}^n$ [22]:

$$\mathbf{y}^{(k+1)} = \begin{cases} \mathbf{0}_n & \|\mathbf{w}^{(k)}\|_2 \leq -s \\ \mathbf{v}^{(k)} & \|\mathbf{w}^{(k)}\|_2 \leq s \\ \left(\frac{1}{2}\right)\left(1 + \frac{s}{\|\mathbf{w}\|_2}\right)[\mathbf{w}^T, \|\mathbf{w}\|_2]^T & \|\mathbf{w}^{(k)}\|_2 \geq |s| \end{cases} \quad (15)$$

The key for hardware-based calculation of $\mathbf{y}^{(k+1)}$ is the $\ell_2$-norm calculation of vector $\mathbf{w}^{(k)}$, which could be performed using peripheral circuits including analog multipliers, summing amplifiers, etc., in the analog domain [17][23]. An alternative method is to convert vector $\mathbf{w}^{(k)}$ to the digital domain and then calculate the $\ell_2$-norm. We will demonstrate in the computational complexity analysis (in Section 3.5) that the overall computational complexity does not increase even if we use only one single ADC/DAC and calculate $\|\mathbf{w}\|_2$ in a sequential manner. After the $\ell_2$-norm calculation, $\mathbf{y}^{(k+1)}$ is calculated based on the comparison results shown in Eqn. (15). This comparison can be implemented in either the analog domain or the digital domain.

### C. Memristor Conductance Matrix Mapping and Elimination of Negative Elements

For using a memristor crossbar to represent the left-hand side matrix in (13), one way is to use Eqn. (5) to map the left-hand side matrix onto the memristor conductance matrix $\mathbf{G}$. However, since the mapping from matrix $\mathbf{C}$ to $\mathbf{G}$ is not a direct one-to-one mapping, it is highly complicated to use Eqn. (5) to perform the mapping. Hence, we adopt a simple and fast approximation: $g_{i,j} = c_{i,j} \cdot g_{max}$, where $g_{max}$ is the maximum conductance among memristors in the memristor crossbar, $c_{i,j}$ represents an element of the left-hand side matrix, and $g_{i,j}$ satisfies: $g_{max} \leq g_{i,j} \leq 0$ [16]. It is proved in [16] that such a simple mapping rule results in negligible inaccuracy. Therefore, the memristor conductance matrix $\mathbf{G}$ is:

$$\mathbf{G} = g_{max} \cdot \begin{bmatrix} \mathbf{I}_{n \times n} & \mathbf{A}^T \\ \mathbf{A} & \mathbf{0}_{m \times m} \end{bmatrix} \quad (16)$$

Based on the matrix mapping results, the solution of the linear system (13) is obtained from:

$$\begin{bmatrix} \mathbf{x} \\ \boldsymbol{\lambda} \end{bmatrix} = \frac{g_s}{g_{max}} \mathbf{V}_I \quad (17)$$

where $\mathbf{V}_I$ is the voltage vector read from the inputs of the memristor crossbar.

Negative elements may exist in matrix $\mathbf{A}$, which is provided by the user and is problem-specific. Since the memristance value cannot be a negative number, effective techniques are necessary to eliminate these negative elements to facilitate memristor crossbar-based implementations. Next, we present an effective method for the elimination of negative coefficients. Consider a linear system $\mathbf{A}\mathbf{x} = \mathbf{b}$, and suppose that $a_{i,j}$ is a negative element in $\mathbf{A}$. The equation in the $i$-th row:

$$a_{i,1}x_1 + \cdots + a_{i,j}x_j + \cdots + a_{i,n}x_n = b_i \quad (18)$$

is equivalent to:

$$\begin{cases} a_{i,1}x_1 + \cdots + 0 \cdot x_j + \cdots + a_{i,n}x_n + (-a_{i,j})z_j = b_i \\ x_j + z_j = 0 \end{cases} \quad (19)$$

Hence, a negative element can be eliminated by setting it to zero and introducing one more row and one more column. Thus, the linear system $\mathbf{A}\mathbf{x} = \mathbf{b}$ can be written as:

$$\begin{bmatrix} a_{1,1} & \ldots & a_{1,j} & \ldots & a_{1,n} & 0 \\ \vdots & \ldots & \ldots & \ldots & \vdots & 0 \\ a_{i,1} & \ldots & 0 & \ldots & a_{i,n} & -a_{i,j} \\ \vdots & \ldots & \ldots & \ldots & \vdots & 0 \\ a_{n,1} & \ldots & a_{n,j} & \ldots & a_{n,n} & 0 \\ 0 & 0 & 1 & 0 & 0 & 1 \end{bmatrix} \begin{bmatrix} x_1 \\ \vdots \\ x_j \\ \vdots \\ x_n \\ z_j \end{bmatrix} = \begin{bmatrix} b_1 \\ \vdots \\ b_j \\ \vdots \\ b_n \\ 0 \end{bmatrix} \quad (20)$$

Similarly, after applying the above technique to Eqn. (13), the linear system can be reformulated as:

$$\begin{bmatrix} I_{n\times n} & A^{T'}_{n\times m} & 0 \text{ or } A^{T''}_{n\times 2k} \\ A'_{m\times n} & 0_{m\times m} & 0 \text{ or } A''_{m\times 2k} \\ A^I_{2k\times n} & A^{TI}_{2k\times m} & 0 \text{ or } I_{2k\times 2k} \end{bmatrix} \begin{bmatrix} x \\ \lambda \\ z \end{bmatrix} = \begin{bmatrix} u^{(k)} \\ b \\ 0 \end{bmatrix} \quad (21)$$

where $A'$ and $A^{T'}$ are obtained by setting the negative elements in $A$ and $A^T$ to zero, respectively. $A''$ and $A^{T''}$ are matrices whose elements are zeros and the absolute values of negative elements in $A$ and $A^T$, respectively. $A^I$ and $A^{TI}$ are matrices consisting of ones and zeros. Locations of ones in $A^I$ and $A^{TI}$, and locations of those absolute values in $A''$ and $A^{T''}$ depend on the locations of negative elements in $A$ and $A^T$. $z$ is a compensation vector of $x$, and will not be further utilized after (21) is solved.

For simplicity, Eqn. (21) can also be written as:

$$M \cdot s = r \quad (22)$$

where $M$ represents the left-hand side matrix in (21), $s$ represents the solution of (21), and $r$ represents the right-hand side vector of (21). Consequently, the memristor conductance matrix of $M$ is set by:

$$G' = g_{max} \cdot M \quad (23)$$

The relation between $s$ and $V_I$ satisfies:

$$s = \frac{g_s}{g_{max}} V_I \quad (24)$$

### D. Detailed Procedure for Solving SOCP Problems Using Memristor Crossbar-Based Framework

The detailed procedure of our proposed memristor crossbar-based framework for solving SOCP problems is summarized as follows:

---

**Detailed Procedure for Solving SOCP Problems Using Memristor Crossbar-Based Framework**

**Input:** Matrix $A$, vectors $b, c, u^{(0)}, \mu^{(0)}$, constant $\varepsilon, \rho, k$.

Output: Vector $x, y$

Initialize $x, y$ with arbitrary elements, $k = 0$.

Construct matrix $M$ and vector $r$ in (22) based on $A$.

Map $M$ to memristor crossbar according to $G' = g_{max} \cdot M$.

do:
1) Solve $x^{(k+1)}$: read the solution $V_I$ from the memristor crossbar, then according to (24), $x^{(k+1)} = \frac{g_s}{g_{max}} V_{I(1:n)}$.
2) Calculate $v^{(k)} \equiv x^{(k+1)} + (1/\rho)\mu^{(k)}$ using summing amplifier, and then construct $w^{(k)}$ and $s$ according to $v^{(k)} = \left[(w^{(k)})^T, s\right]^T$.
3) Calculate $y^{(k+1)}$ in (15) by calculating $w^{(k)}$ with peripheral circuits.
4) Update $\mu^{(k+1)}$ using summing amplifier according to (7c).
5) Update $u^{(k+1)}$ using summing amplifier: $u^{(k+1)} = y^{(k+1)} - (1/\rho)(\mu^{(k+1)} + c)$.
6) $k = k + 1$.

While $x^{(k+1)} - x^{(k)} > \varepsilon$

Return $x, c^T x$.

---

### E. Computational Complexity Analysis

The algorithm-hardware co-optimization of the memristor-based framework proposed in this paper is an iterative solution framework. In each iteration, the complexity of solving Eqn. (21) with the memristor crossbar is O(1) and that of calculating $y^{(k+1)}$ with peripheral circuits is O($N$). Hence the framework presents an overall solution complexity of pseudo-O($N$), or O($MN$) if $M$ represents the number of iterations, which is a significant improvement compared with the complexity of the state-of-the-art software-based solution namely O($N^{3.5}$) - O($N^4$). For various scales of optimization problems, the speed improvement compared with software-based solvers can be as high as $10^4$-$10^5$ and the energy efficiency improvement can be even more significant.

The previous discussion only accounts for the solution complexity. The complexity of initialization of the matrix in the memristor crossbar is O($N^2$), or lower for sparse matrices which are very common in (large-scale) optimization problems.

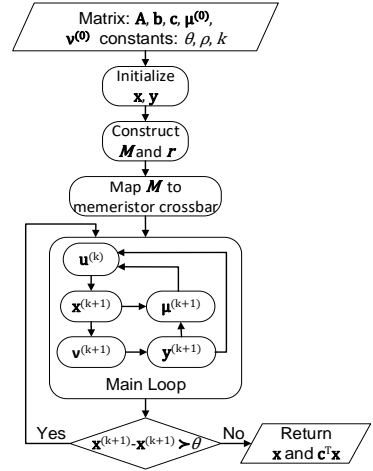

Fig. 2. Flow Diagram of the proposed algorithm for solving SOCP problems

### F. Memristor-Based Framework for Solving Homogeneous QCQP Problems

Consider the homogeneous QCQP problem expressed in (4). According to the eigenvalue decomposition of $P_i$, we have $Q_i$ such that $P_i = Q_i^T Q_i$. Upon defining:

$$z_i = \begin{bmatrix} Q_i x \\ \sqrt{r_i} \end{bmatrix} \in \mathbb{R}^{n+1},$$

problem (4) can be expressed as a convex program with second-order cone constraints:

$$\begin{aligned} \text{minimize} \quad & x^T P_0 x \\ \text{subject to:} \quad & \|[z_i]_{1:n}\|_2 \leq [z_i]_{n+1} \; (i=1,\ldots,m) \\ & z_i - C_i x = d_i \; (i=1,\ldots,m) \\ & Ax = b, \end{aligned} \quad (25)$$

where the optimization variables are $x$ and $z_i$, and $C_i = [Q_i^T, 0]^T$, $d_i = [0^T, \sqrt{r_i}]^T$.

After applying the ADMM technique to problem (25), it can be written as:

$$\begin{aligned} \text{minimize} \quad & x^T P_0 x + \sum_{i=1}^m I(z_i) \\ \text{subject to:} \quad & z_i - C_i x = d_i, i=1,2,\ldots,m, \\ & Ax = b, \end{aligned} \quad (26)$$

where $I(z_i)$ is an indicator functions given by

$$I(z_i) = \begin{cases} 0, & \text{if } \|[z_i]_{1:n}\|_2 \leq [z_i]_{n+1} \\ \infty, & \text{otherwise} \end{cases} \quad (27)$$

The augmented Lagrangian of (26) is given by

$$\mathcal{L}(x, \{z_i\}, \{u_i\}, v) = x^T P_0 x + \sum_{i=1}^m I(z_i) + \sum_{i=1}^m u_i^T(z_i - C_i x - d_i) + \frac{\rho}{2}\|z_i - C_i x - d_i\|_2^2 + v^T(Ax - b) + \frac{\rho}{2}\|Ax - b\|_2^2, \quad (28)$$

where $u_i$ and $v$ are Lagrangian multipliers, and their values at $(t+1)$-th iteration are given by:

$$u_i^{(t+1)} = u_i^{(t)} + \rho(z_i^{(t+1)} - C_i x^{(t+1)} - d_i) \quad (29)$$
$$v^{(t+1)} = v^{(t)} + \rho(Ax^{(t+1)} - b). \quad (30)$$

Solving problem (28) yields the x-minimization step and the z-minimization step:

- x-minimization step:
$$\text{minimize} \quad x^T P_0 x + \frac{\rho}{2}\sum_{i=1}^{m}\left\|z_i^{(t)} - C_i x - d_i + \frac{1}{\rho}u_i^{(t)}\right\|_2^2 + \frac{\rho}{2}\left\|Ax - b + \frac{1}{\rho}v^{(t)}\right\|_2^2. \quad (31)$$

Let $g_i^{(t)} = z_i^{(t)} - d_i + \frac{1}{\rho}u_i^{(t)}$, and $h^{(t)} = b - \frac{1}{\rho}v^{(t)}$, the solution of (31) yields a system of linear equations:

$$(2P_0 + \rho\sum_{i=1}^{m} C_i^T C_i + \rho A^T A)x = \rho\sum_{i=1}^{m} C_i^T g_i + \rho A^T h. \quad (32)$$

where $C_i^T C_i = P_i$.

- z-minimization step:
$$\begin{aligned}\text{minimize} \quad &\|z_i - L\|_2^2 \\ \text{subject to:} \quad &\|[z_i]_{1:n}\|_2 \leq [z_i]_{n+1}\end{aligned} \quad (33)$$

where $L = C_i x^{(t+1)} + d_i - \frac{1}{\rho}u_i^{(t)}$.

ADMM terminates when the following two conditions are satisfied:

$$\left\|x^{(t+1)} - x^{(t)}\right\|_2 + \sum_{i=1}^{m}\left\|z_i^{(t+1)} - z_i^{(t)}\right\|_2 \leq \epsilon,$$

and $\left\|Ax^{(t+1)} - b\right\|_2 + \sum_{i=1}^{m}\left\|z_i^{(t+1)} - C_i x^{(t+1)} - d_i\right\|_2 \leq \epsilon.$

Eqn. (32) and Eqn. (33) are respectively similar to Eqn. (13) and Eqn. (11) for solving SOCP problems. Hence our proposed memristor-based framework can also be utilized to solve the homogeneous QCQP problems.

## IV. Experiments And Analysis

### A. Estimation of Power and Performance Improvement

Our estimation of power and performance improvement is based on accurate memristor modeling work from [29], demonstrating significant improvement in speed and energy efficiency of memristor crossbar based solution framework for SOCP/homogeneous QCQP problems. The estimated time for solving SOCP/homogeneous QCQP problems is around or less than 400μs if the number of variables is 1024. This estimation is based on (i) actual simulation results indicating that it generally takes 500-800 iterations for convergence, and (ii) the amount of analog or digital computation in one iteration (the most time-consuming computation is calculating $y^{(k+1)}$) can be performed in at most N clock cycles (where N is the number of variables. This is a very conservative estimate.). Please note that our procedure does not need to update matrix M in the memristor crossbar during iterations. A maximum of $1.57 \times 10^5$ X estimated improvement in speed is achieved compared with the CVX tool executed on an Intel I7 server. This significant improvement comes from the reduction in computational complexity and the speedup due to dedicated hardware implementations. The maximum amount of energy reduction is $1.32 \times 10^7$X in this case, which is even more significant than the speedup, because of the low power consumption of the dedicated hardware of memristor crossbar and peripheral circuits.

### B. Experiments to Study the Effect of Process Variations

We develop a simulation and evaluation framework for using memristor crossbars to solve SOCP and homogeneous QCQP problems that accounts for the effect of process variations. The developed simulation framework simulates the memristor crossbar-based iterative solution framework, and in each iteration it calculates outputs of the memristor crossbar from inputs by solving KCL/KVL equations. Element writing inaccuracies, random process variations, and other variations can be accounted for in the simulation framework and their impact on the outcomes can be evaluated.

All input matrices are randomly generated using *randn* and *sprandn* functions provided by Matlab. These inputs are first sent to the CVX tool, which is a well-known and widely recognized software solver for convex optimization. After the randomly generated problems are verified to be feasible and bounded, our solver is then utilized to solve the problems.

Generally, the constraint matrix of an SOCP or homogeneous QCQP problem is sparse. Thus, in our experiments, the constraint matrix $A$ (corresponding to the constraints $Ax = b$) is sparse with respective density of 0.1 (i.e., only 10% elements in $A$ are nonzero) and 0.2. The size of the optimization vector $x$ is $2^i$, $i = 4, 5, 6, 7, 8, 9, 10$. Although the size of a single memristor crossbar is limited to 1024 at most, multiple crossbars can be organized in the structure of *network-on-chip* (*NoC*) to perform larger-scale computations [30]. Additionally, due to the unavoidable variations introduced by the matrix elements writing errors, manufacturing process, and temperature changes, etc., it is necessary to test the accuracy of our proposed tool and how variation-tolerant it is. Thus, for each individual problem, three scenarios with process variations of 0%, 5% and 10% are simulated, and each scenario is simulated 200 times.

### C. Experimental Results and Analysis

Fig. 3-(a) and Fig. 3-(b) show the error rates and failure rates when solving the SOCP problem under different conditions (different problem/matrix size, process variations, etc.), using our developed simulation/evaluation framework. Error here means the difference between the result solved by our framework and the result solved by CVX, and failure here means the result does not converge to the optimal value obtained by CVX. We find that under the ideal condition, i.e., no variation is introduced, the optimal values are the same as the results calculated by CVX and no failure is found. And it is obvious that higher variation level will result in lower accuracy and higher failure rates.

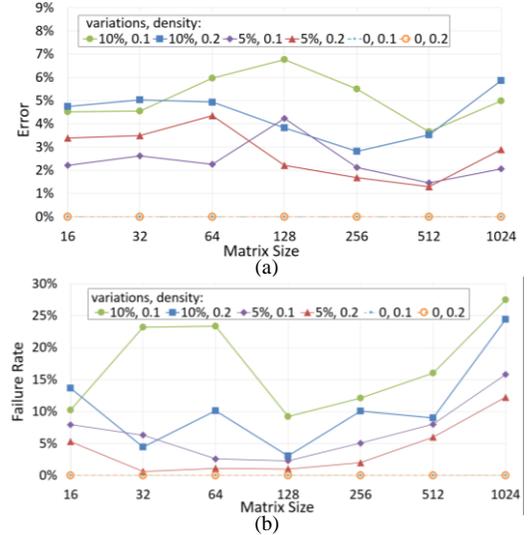

Fig. 3. Simulation results for solving SOCP problems (a) Error vs Matrix Size (b) Failure vs Matrix Size

From the point of view of matrix density, the cases with density of 0.1 will result in higher failure rates than the cases with density of 0.2, namely, the denser the input matrix $A$ is, the more reliability our hardware-based solution framework can guarantee. From the point of view of matrix size, the success rates decrease with the increase in the size of matrix $A$. Basically, our algorithm can achieve high accuracy (95%) and success rates (85%) when the process variations are restricted to below 5%.

Fig. 4-(a) depicts the error rates while solving homogeneous QCQP problems when $A$ has different sparsity levels and variation levels, meanwhile a strict ADMM tolerance is set (i.e., $\varepsilon$ is less than $1.0 \times 10^{-4}$). This strict ADMM tolerance can result in high time complexity, i.e., solving each problem usually iterates $1/\varepsilon$ times or so. Under such strict tolerance, the accuracy is relatively low when the problem size is less than 128, but the problems of large-size converge very well. Therefore, we can conclude that our framework is more appropriate for large-sized homogeneous QCQP problems. Due to the fact that error rates are less than 3%, there is a lot of room for tradeoffs between accuracy and time complexity. Thus, we relax the ADMM tolerance to $1.0 \times 10^{-3}$ to reduce the number of iterations, and Fig. 4-

(b) displays the error rates with more relaxed tolerance level. We find that even though we significantly reduce the number of iterations, up to 96% accuracy can be obtained. Unlike the results for SOCP problems, the impact of process variations is quite limited here and not a single failure is found during the experiments. Because the source of accuracy loss is that variations somehow change the original problem (the objective function and constraints) by changing the mapped matrix, considering Eqn. (32), the effects of variations added to the memristor crossbar is reduced by a factor of $\rho$. Hence when $\rho$ is large, the effect is small.

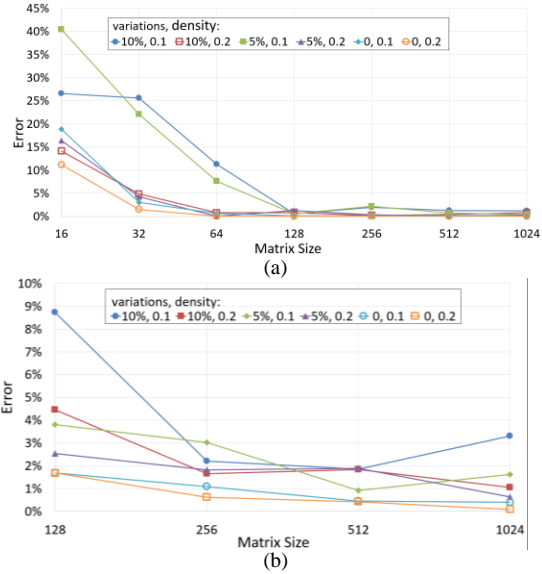

Fig. 4. Simulation results for solving homogeneous QCQP problems (a) errors with strict ADMM tolerance (b) errors with relaxer ADMM tolerance

Considering all simulation cases, we conclude that our framework is a reliable tool for solving SOCP problems if the process variations can be restricted to within 5%, and the framework is suitable for different matrix sizes. For homogeneous QCQP problems, our solver is more appropriate for large-sized problems, and it is more robust, accurate and variation-tolerant than for solving SOCP problems.

## V. Conclusion

This paper introduced the memristor device and its crossbar structure, to solve SOCP as well as homogeneous QCQP problems. To use the memristor crossbar to solve the above two problems, the presented framework in this paper applied ADMM to decompose the two problems into linear systems, so that the merit of memristor crossbar for solving a set of linear equations in O(1) time complexity, can be sufficiently utilized. The overall time complexity of solving SOCP and homogeneous QCQP problems are both pseudo-O($N$). Our experiments demonstrated that our proposed algorithm can achieve higher than 94% accuracy when solving SOCP problems and higher than 96% accuracy when solving homogeneous QCQP problems.